\documentclass[conference]{IEEEtran}

\renewcommand\IEEEkeywordsname{Index Terms}
\usepackage{graphicx}
\usepackage{subcaption}
\usepackage{tikz,xparse}
\usetikzlibrary{dsp,chains}
\usetikzlibrary{matrix}
\usepackage{mathptmx}
\usepackage{verbatim}
\usepackage{calc}
\usepackage{ifthen}
\usepackage{xifthen}
\usepackage{cancel}
\usepackage{bm}
\usepackage{verbatim}
\usepackage{multirow}
\usepackage{cite}

\usepackage[nolist]{acronym} 
\usepackage{pgfplots}
\usetikzlibrary{arrows,shapes,graphs,graphs.standard,quotes}
\pgfplotsset{compat=newest}
\usetikzlibrary{matrix}

\usepackage[hyphens]{url}

\usepackage[bookmarks=false]{hyperref}
\usepackage{units}
\usepackage{amsmath, amsbsy, amssymb, latexsym }
\hypersetup{bookmarksdepth=-2}
\usepackage{comment}
\usepackage[utf8]{inputenc}
\usepackage{xcolor}
\usepackage{enumitem}
\usepackage{algorithm}
\usepackage{algpseudocode}

\usepackage{etoolbox} 

\tikzset{>=latex}

\algnewcommand\algorithmicforeach{\textbf{for each}}
\algdef{S}[FOR]{ForEach}[1]{\algorithmicforeach\ #1\ \algorithmicdo}

\algnewcommand\algorithmicswitch{\textbf{switch}}
\algnewcommand\algorithmiccase{\textbf{case}}
\algnewcommand\algorithmicassert{\texttt{assert}}
\algnewcommand\Assert[1]{\State \algorithmicassert(#1)}%
\algdef{SE}[SWITCH]{Switch}{EndSwitch}[1]{\algorithmicswitch\ #1\ \algorithmicdo}{\algorithmicend\ \algorithmicswitch}%
\algdef{SE}[CASE]{Case}{EndCase}[1]{\algorithmiccase\ #1}{\algorithmicend\ \algorithmiccase}%
\algtext*{EndCase}%

\definecolor{mittelblau}{RGB}{0, 126, 198}
\definecolor{violettblau}{cmyk}{0.9, 0.6, 0, 0}
\definecolor{rot}{RGB}{238, 28 35}
\definecolor{apfelgruen}{RGB}{140, 198, 62}
\definecolor{gelb}{RGB}{255, 229, 0}
\definecolor{orange}{RGB}{244, 111, 33}
\definecolor{pink}{RGB}{237, 0, 140}
\definecolor{lila}{RGB}{128, 10, 145}
\definecolor{hellgrau}{RGB}{224, 224, 224}
\definecolor{mittelgrau}{RGB}{128, 128, 128}
\definecolor{dunkelgrau}{RGB}{80,80,80}
\definecolor{anthrazit}{RGB}{19, 31, 31}
\definecolor{darkgreen}{RGB}{34,139,34}
\colorlet{Mycolor1}{green!10!orange!90!}
    
\tikzset{
        vnds/.style={ shape=circle, fill=black, draw, inner sep=0pt,minimum size=5pt},
        cnds/.style={ shape=rectangle, fill=white, draw, inner sep=0pt,minimum size=5pt}, 
        vndGs/.style={ shape=circle, fill=green, draw, inner sep=0pt,minimum size=5pt},
        vndGsg/.style={ shape=circle, draw=black, fill=green, draw, inner sep=0.05pt,minimum size=5pt},
        vndRs/.style={ shape=circle, fill=red, draw, inner sep=0pt,minimum size=5pt},     
        vndRsr/.style={ shape=circle, draw=red, fill=red, draw, inner sep=0pt,minimum size=5pt},     
        vndsc/.style={ shape=circle, fill=black, draw, inner sep=0pt,minimum size=7.5pt},
        cndsc/.style={ shape=rectangle, fill=white, draw, inner sep=0pt,minimum size=7.5pt}, 
        vndGsc/.style={ shape=circle, fill=green, draw, inner sep=0pt,minimum size=7.5pt},
        vndRsc/.style={ shape=circle, fill=red, draw, inner sep=0pt,minimum size=7.5pt}
}
\def\angleRot{0} \def\vndPos{left} \def\cndPos{right}

\IEEEoverridecommandlockouts

\begin{document}
	
	\begin{NoHyper}
		\title{Optimizing Polar Codes Compatible with Off-the-Shelf LDPC Decoders}

		\author{\IEEEauthorblockN{Moustafa Ebada, Ahmed Elkelesh and Stephan ten Brink} \thanks{This work has been supported by DFG, Germany, under grant BR 3205/5-1.}
			\IEEEauthorblockA{
				Institute of Telecommunications, Pfaffenwaldring 47, University of  Stuttgart, 70569 Stuttgart, Germany 
				\\\{ebada,elkelesh,tenbrink\}@inue.uni-stuttgart.de
			}
		}

		\makeatletter
		\patchcmd{\@maketitle}
		{\addvspace{0.5\baselineskip}\egroup}
		{\addvspace{-0.6\baselineskip}\egroup}
		{}
		{}
		\makeatother

		\maketitle
		
		\begin{acronym}
			\acro{ECC}{error-correcting code}
			\acro{HDD}{hard decision decoding}
			\acro{SDD}{soft decision decoding}
			\acro{ML}{maximum likelihood}
			\acro{GPU}{graphical processing unit}
			\acro{BP}{belief propagation}
			\acro{BPL}{belief propagation list}
			\acro{LDPC}{low-density parity-check}
			\acro{HDPC}{high density parity check}
			\acro{BER}{bit error rate}
			\acro{SNR}{signal-to-noise-ratio}
			\acro{BPSK}{binary phase shift keying}
			\acro{AWGN}{additive white Gaussian noise}
			\acro{MSE}{mean squared error}
			\acro{LLR}{Log-likelihood ratio}
			\acro{MAP}{maximum a posteriori}
			\acro{NE}{normalized error}
			\acro{BLER}{block error rate}
			\acro{PE}{processing elements}
			\acro{SCL}{successive cancellation list}
			\acro{SC}{successive cancellation}
			\acro{BI-DMC}{Binary Input Discrete Memoryless Channel}
			\acro{CRC}{cyclic redundancy check}
			\acro{CA-SCL}{CRC-aided successive cancellation list}
			\acro{BEC}{Binary Erasure Channel}
			\acro{BSC}{Binary Symmetric Channel}
			\acro{BCH}{Bose-Chaudhuri-Hocquenghem}
			\acro{RM}{Reed--Muller}
			\acro{RS}{Reed-Solomon}
			\acro{SISO}{soft-in/soft-out}
			\acro{PSCL}{partitioned successive cancellation list}
			\acro{3GPP}{$3^{\text{rd}}$ generation partnership project}
			\acro{eMBB}{enhanced Mobile Broadband}
			\acro{PCC}{parity-check concatenated}
			\acro{CA-polar codes}{CRC-aided polar codes}
			\acro{CN}{check nodes}
			\acro{PC}{parity-check}
			\acro{GenAlg}{Genetic Algorithm}
			\acro{AI}{Artificial Intelligence}
			\acro{MC}{Monte Carlo}
			\acro{CSI}{Channel State Information}
			\acro{PSCL}{partitioned successive cancellation list}
			\acro{5G-NR}{5G new radio}
			\acro{5G}{$5^{th}$ generation mobile communication}
			\acro{mMTC}{massive machine-type communications}
			\acro{URLLC}{ultra-reliable low-latency communications
			}

		\end{acronym}
		
		\begin{abstract}
			
			Previous work showed
			that polar codes can be decoded using off-the-shelf LDPC decoders by 
			imposing special constraints on the
			LDPC code structure, which, however, resulted in some performance 
			degradation. 
			In this paper we show that this loss can be mitigated;
			in particular, we demonstrate
			how the gap  between LDPC-style decoding
			and Arıkan's Belief Propagation (BP) decoding of polar
			codes can be closed by taking into account the underlying 
			graph structure of the LDPC decoder while jointly designing the polar code and 
			the parity-check matrix of the corresponding LDPC-like code. The resulting polar codes under conventional
			LDPC-style decoding are shown to have similar error-rate performance
			when compared to some well-known and standardized LDPC codes. Moreover, we obtain performance gains in the high
			SNR region.
			
			
		\end{abstract}
		\acresetall
		\vspace{-0.2cm}
		\section{Introduction}

		
		Polar codes, proposed in  \cite{ArikanMain}, are known to be the first class of error-correcting codes that is theoretically proven to be asymptotically capacity-achieving under \ac{SC} decoding. In the ongoing \ac{5G} standardization process, the \ac{3GPP} group has decided to deploy polar codes for the uplink and downlink control channel of the \ac{eMBB} service \cite{Huawei}, while discussion about deploying polar codes for \ac{mMTC} and \ac{URLLC} is still ongoing. Therefore, developing practical decoders that satisfy the requirements defined by \ac{5G-NR} has been progressively very intense, active and demanding.
		One of the issues targeted by \ac{5G-NR} is that it should have latency of less than 1 ms compared to 10 ms of 4G systems \cite{HuaweiWhitePaper}.  This latency reduction is required to enable the newly targeted applications of 5G systems such as augmented reality and 3D video rendering.

		The fact that the \ac{SC} decoder of polar codes suffers from a weak error-rate 
		performance for finite-length codes has urged the development of further 
		decoders. In \cite{talvardyList}, the \ac{SC} decoder was extended to a 
		\ac{SCL} decoder by applying the list decoding scheme on the plain \ac{SC} 
		decoder leading to an improved error-rate performance, close to the \ac{ML} 
		bound for sufficiently large list sizes, on the cost of increased complexity. 
		It was observed that in order not to miss the correct 
		codeword in the list, a high-rate \ac{CRC} could be concatenated with the 
		polar code and used while decoding, or just at the very end, as a path tag 
		(i.e., distributed or non-distributed \ac{CRC}, respectively), explaining the 
		name \ac{CA-SCL}. This concatenation leads to significant performance gains over state-of-the-art \ac{LDPC} and turbo codes. The concatenation of a polar code 
		with a high-rate parity-check code was likewise introduced in 
		\cite{trifonovDynFr} and \cite{PCC} and shown to be of similar error-rate 
		performance as of the \ac{CRC}-aided polar codes, with more simplicity and 
		flexibility. The resultant codes enjoy better weight spectrum properties 
		\cite{AdaptiveList} which is regarded as one reason for the improved performance. The 
		performance of polar codes (i.e., both \emph{plain} and 
		\emph{\ac{CRC}-aided})  were further pushed forward by tailoring the designed 
		polar code to both specifics of the code and decoder 
		\cite{GenAlg_Journal_IEEE} at the cost of reduced code flexibility and increased code design complexity 
		(i.e., offline complexity).
		
		A major drawback of \ac{SC}-based decoders is the high decoding latency and, thus, 
		their low throughput due to the sequential decoding manner. To combat that,  
		various \ac{SC}-based decoding schemes were proposed, e.g., 
		\cite{fastListDecoder} and \cite{fastSCF}, mostly based on identifying specific 
		node patterns and efficiently utilizing them to speed up the decoding process. 
		Besides, for an \ac{SCL} decoder, there is an increased implementation 
		complexity (e.g., hardware complexity, memory requirements, etc.) proportional 
		to the list size deployed.  For that, more simplified variations were 
		introduced, e.g., \cite{AdaptiveList} and \cite{PSCL_Journal}, limiting the 
		increase in complexity, with only minimal performance loss in some cases. Furthermore, 
		these classes of decoders do not provide soft-in/soft-out information 
		processing which limits their usage in iterative detection and 
		decoding schemes.
		
				
		Usually a standard does not define a certain decoding technique and, thus, improved decoding algorithms (in terms of performance, complexity or throughput) may result in a competitive advantage.
		There exists another family of polar decoders, namely iterative decoders, which do not have the aforementioned problems. This class of decoders could be potentially suitable for high data rate applications (i.e., in particular from an implementation point of view). The first iterative decoder proposed was the \ac{BP} decoder \cite{ArikanBP} which was conducted on the factor graph that corresponds to the polarization matrix of the code. 
		However, this decoder has an error-rate performance which is inferior to that 
		of the \ac{CA-SCL} decoder. One further step was made in 
		\cite{elkelesh2018belief}, where a \ac{BPL} decoder was introduced, with an 
		improved error-rate performance compared to any other 
		known-thus-far iterative decoder, however,  while still being inferior to  
		\ac{CA-SCL} in terms of error-rate performance; this can be attributed to the 
		\ac{CRC} incompatibility and the non-optimal code design in the case of the 
		\ac{BPL} decoder.
		
		In this work, we use the \ac{GenAlg} to find (design) better polar codes (i.e., information/frozen sets) tailored to LDPC-style BP decoding.
		Also, we investigate the effect of changing the information/frozen set on the graphical representation of the underlying polar code which is used to run the BP algorithm.
					\vspace{-0.1cm}
		\section{Polar Codes Interpreted as LDPC Codes} \label{sec:con}
					\vspace{-0.1cm}
		Polar codes  can be interpreted as \ac{LDPC} codes with an underlying sparse 
		Tanner graph \cite{sparseBP}. This is illustrated in Fig.~\ref{fig:bip}, Arıkan's original factor graph on the left and its corresponding 
		bipartite factor graph, i.e., $\mathbf{H}_\text{sparse}$, on the right.
		With some basic pruning techniques, the size of the bipartite graph can be significantly 
		reduced and made practical to use. This means that \emph{any} polar code (e.g., the 
		polar code considered in the \ac{5G} standard) can be surprisingly decoded using 
		\emph{conventional, off-the-shelf} \ac{LDPC} decoders, e.g., based on the 
		sum-product algorithm (SPA), given that we impose some constraints on the 
		underlying \ac{LDPC} code structure. On the one hand, this has the great 
		advantage of re-using the existing hardware implementations of \ac{LDPC} 
		decoders, in addition to making use of the available systematic \ac{LDPC} 
		analysis and design tools; and enjoying all complexity, memory requirements 
		and latency being significantly reduced. On the other hand, this way of 
		interpreting/decoding the polar code has the drawback of degraded error-rate 
		performance when compared to the conventional \ac{BP} decoder \cite[Fig.~1]{sparseBP}. The decoding complexity of the LDPC-style decoder was 
		significantly improved in \cite{yairSparseDL} by means of deep learning with 
		further enhanced throughput. 
		
		\begin{figure}[t]
			\vspace{-0.5cm}
			\resizebox{0.95\columnwidth}{!}{
				
				\begin{subfigure}{0.5\columnwidth}
					\resizebox{\columnwidth}{!}{\tikzset{text=black, font={\fontsize{10pt}{10}\selectfont}}

\let\pgfmathMod=\pgfmathmod\relax


\def\n{3}  \def\N{8}
\begin{tikzpicture}
\tikzset{edge/.style = {-, thick}}

\tikzset{h1/.style={preaction={
draw,yellow,-,
double=yellow,
double distance=4\pgflinewidth,
}}}

\tikzset{h2/.style={preaction={
draw,green,-,
double=green,
double distance=4\pgflinewidth,
}}}
          \pgfmathtruncatemacro{\nn}{(\n + 1)}
          
\foreach \i in {1,...,\N} {
\foreach \j in {1} {
	   \ifthenelse{\i=1 \OR \i=2 \OR \i=3 \OR \i=5}
	{\node [vndRsc, label=above:{$v_{\i\j}$}] (v\i\j) at (2*\j ,13.5 -1.5*\i) 
	{};	
	}
	{\node [vndsc, label=above:{$v_{\i\j}$}] (v\i\j) at (2*\j ,13.5 -1.5*\i) 
	{};	
	}
   }}

    \foreach \i in {1,...,\N} {
\foreach \j in {1,...,\n} {
   \ifthenelse{\j>1}
		{\node [vndsc, label=above:{$v_{\i\j}$}] (v\i\j) at (2*\j ,13.5 
		-1.5*\i) 
		{};
       		\node [cndsc, label=above:{$c_{\i\j}$}] (c\i\j) at (2*\j+1 ,13.5 
       		-1.5*\i) 
       		{};}      
       		{\node [cndsc, label=above:{$c_{\i\j}$}] (c\i\j) at (2*\j +1,13.5 
       		-1.5*\i) 
       		{};} 		
   }}

\foreach \i in {1,...,\N} {
\foreach \j in {4} {
	\node [vndGsc, label=above:{$v_{\i\j}$}] (v\i\j) at (2*\j ,13.5 -1.5*\i) 
	{};	
   }}

    \foreach \i in {1,...,\N} {
       \foreach \j in {1,...,\n} {
        		\draw (v\i\j)--(c\i\j);

   }}
   
      \foreach \i in {1,...,\n} {
                \pgfmathtruncatemacro{\ii}{(\i + 1)}
      \foreach \j in {1,...,\N} {
         		\draw (c\j\i)--(v\j\ii);

   }}

   \foreach \i [count=\xi] in {1,2,3} {
  \ifthenelse{\xi=1}{
  \draw (c1\i)--(v5\i);
   \draw (c2\i)--(v6\i);
   \draw (c3\i)--(v7\i);
   \draw (c4\i)--(v8\i);
   } {       \ifthenelse{\xi=2}{
     \draw (c1\i)--(v3\i);
   \draw (c2\i)--(v4\i);
   \draw (c5\i)--(v7\i);
   \draw (c6\i)--(v8\i);
   }{
     \draw (c1\i)--(v2\i);
   \draw (c3\i)--(v4\i);
   \draw (c5\i)--(v6\i);
   \draw (c7\i)--(v8\i);
     }
     }
}
\end{tikzpicture}} 
					\label{fig:conv}
				\end{subfigure}
								$\Leftrightarrow$
				\begin{subfigure}{0.2\columnwidth}
					\resizebox{\columnwidth}{!}{\tikzset{text=black, font={\fontsize{10pt}{10}\selectfont}}
\let\pgfmathMod=\pgfmathmod\relax
\definecolor{lightgrey}{gray}{0.9}

\begin{tikzpicture}[rotate=\angleRot]
\tikzset{edge/.style = {-, thick}}

\tikzset{h1/.style={preaction={draw,yellow,-,double=yellow,double 
distance=4\pgflinewidth,}}}

\tikzset{h2/.style={preaction={draw,green,-,double=green,double 
distance=4\pgflinewidth,}}}

\node [cnds, label=\cndPos:{$c_{11}$}](c11) at(3,12) {};
\node [cnds, label=\cndPos:{$c_{21}$}](c21) at(3,11.5) {};
\node [cnds, label=\cndPos:{$c_{31}$}](c31) at(3,11) {};
\node [cnds, label=\cndPos:{$c_{41}$}](c41) at(3,10.5) {};
\node [cnds, label=\cndPos:{$c_{51}$}](c51) at(3,10) {};
\node [cnds, label=\cndPos:{$c_{61}$}](c61) at(3,9.5) {};
\node [cnds, label=\cndPos:{$c_{71}$}](c71) at(3,9) {};
\node [cnds, label=\cndPos:{$c_{81}$}](c81) at(3,8.5) {};
\node [cnds, label=\cndPos:{$c_{12}$}](c12) at(3,8) {};
\node [cnds, label=\cndPos:{$c_{22}$}](c22) at(3,7.5) {};
\node [cnds, label=\cndPos:{$c_{32}$}](c32) at(3,7) {};
\node [cnds, label=\cndPos:{$c_{42}$}](c42) at(3,6.5) {};
\node [cnds, label=\cndPos:{$c_{52}$}](c52) at(3,6) {};
\node [cnds, label=\cndPos:{$c_{62}$}](c62) at(3,5.5) {};
\node [cnds, label=\cndPos:{$c_{72}$}](c72) at(3,5) {};
\node [cnds, label=\cndPos:{$c_{82}$}](c82) at(3,4.5) {};
\node [cnds, label=\cndPos:{$c_{13}$}](c13) at(3,4) {};
\node [cnds, label=\cndPos:{$c_{23}$}](c23) at(3,3.5) {};
\node [cnds, label=\cndPos:{$c_{33}$}](c33) at(3,3) {};
\node [cnds, label=\cndPos:{$c_{43}$}](c43) at(3,2.5) {};
\node [cnds, label=\cndPos:{$c_{53}$}](c53) at(3,2) {};
\node [cnds, label=\cndPos:{$c_{63}$}](c63) at(3,1.5) {};
\node [cnds, label=\cndPos:{$c_{73}$}](c73) at(3,1) {};
\node [cnds, label=\cndPos:{$c_{83}$}](c83) at(3,0.5) {};

\node [vndRsr, label=\vndPos:{$v_{11}$}](v11) at(1,14) {};
\node [vndRsr, label=\vndPos:{$v_{21}$}](v21) at(1,13.5) {};
\node [vndRsr, label=\vndPos:{$v_{31}$}](v31) at(1,13) {};
\node [vnds, label=\vndPos:{$v_{41}$}](v41) at(1,12.5) {};
\node [vndRsr, label=\vndPos:{$v_{51}$}](v51) at(1,12) {};
\node [vnds, label=\vndPos:{$v_{61}$}](v61) at(1,11.5) {};
\node [vnds, label=\vndPos:{$v_{71}$}](v71) at(1,11) {};
\node [vnds, label=\vndPos:{$v_{81}$}](v81) at(1,10.5) {};
\node [vnds, label=\vndPos:{$v_{12}$}](v12) at(1,10) {};
\node [vnds, label=\vndPos:{$v_{22}$}](v22) at(1,9.5) {};
\node [vnds, label=\vndPos:{$v_{32}$}](v32) at(1,9) {};
\node [vnds, label=\vndPos:{$v_{42}$}](v42) at(1,8.5) {};
\node [vnds, label=\vndPos:{$v_{52}$}](v52) at(1,8) {};
\node [vnds, label=\vndPos:{$v_{62}$}](v62) at(1,7.5) {};
\node [vnds, label=\vndPos:{$v_{72}$}](v72) at(1,7) {};
\node [vnds, label=\vndPos:{$v_{82}$}](v82) at(1,6.5) {};
\node [vnds, label=\vndPos:{$v_{13}$}](v13) at(1,6) {};
\node [vnds, label=\vndPos:{$v_{23}$}](v23) at(1,5.5) {};
\node [vnds, label=\vndPos:{$v_{33}$}](v33) at(1,5) {};
\node [vnds, label=\vndPos:{$v_{43}$}](v43) at(1,4.5) {};
\node [vnds, label=\vndPos:{$v_{53}$}](v53) at(1,4) {};
\node [vnds, label=\vndPos:{$v_{63}$}](v63) at(1,3.5) {};
\node [vnds, label=\vndPos:{$v_{73}$}](v73) at(1,3) {};
\node [vnds, label=\vndPos:{$v_{83}$}](v83) at(1,2.5) {};
\node [vndGsg, label=\vndPos:{$v_{14}$}](v14) at(1,2) {};
\node [vndGsg, label=\vndPos:{$v_{24}$}](v24) at(1,1.5) {};
\node [vndGsg, label=\vndPos:{$v_{34}$}](v34) at(1,1) {};
\node [vndGsg, label=\vndPos:{$v_{44}$}](v44) at(1,0.5) {};
\node [vndGsg, label=\vndPos:{$v_{54}$}](v54) at(1,0) {};
\node [vndGsg, label=\vndPos:{$v_{64}$}](v64) at(1,-0.5) {};
\node [vndGsg, label=\vndPos:{$v_{74}$}](v74) at(1,-1) {};
\node [vndGsg, label=\vndPos:{$v_{84}$}](v84) at(1,-1.5) {};

\draw (c11)--(v12);
\draw (c21)--(v22);
\draw (c31)--(v32);
\draw (c41)--(v42);
\draw (c51)--(v52);
\draw (c61)--(v62);
\draw (c71)--(v72);
\draw (c81)--(v82);
\draw (c12)--(v13);
\draw (c22)--(v23);
\draw (c32)--(v33);
\draw (c42)--(v43);
\draw (c52)--(v53);
\draw (c62)--(v63);
\draw (c72)--(v73);
\draw (c82)--(v83);
\draw (c13)--(v14);
\draw (c23)--(v24);
\draw (c33)--(v34);
\draw (c43)--(v44);
\draw (c53)--(v54);
\draw (c63)--(v64);
\draw (c73)--(v74);
\draw (c83)--(v84);

\draw (v11)--(c11);
\draw (v12)--(c12);
\draw (v13)--(c13);
\draw (v21)--(c21);
\draw (v22)--(c22);
\draw (v23)--(c23);
\draw (v31)--(c31);
\draw (v32)--(c32);
\draw (v33)--(c33);
\draw (v41)--(c41);
\draw (v42)--(c42);
\draw (v43)--(c43);
\draw (v51)--(c51);
\draw (v52)--(c52);
\draw (v53)--(c53);
\draw (v61)--(c61);
\draw (v62)--(c62);
\draw (v63)--(c63);
\draw (v71)--(c71);
\draw (v72)--(c72);
\draw (v73)--(c73);
\draw (v81)--(c81);
\draw (v82)--(c82);
\draw (v83)--(c83);

\draw (c11)--(v51);
\draw (c21)--(v61);
\draw (c31)--(v71);
\draw (c41)--(v81);
\draw (c12)--(v32);
\draw (c22)--(v42);
\draw (c52)--(v72);
\draw (c62)--(v82);
\draw (c13)--(v23);
\draw (c33)--(v43);
\draw (c53)--(v63);
\draw (c73)--(v83);

\end{tikzpicture}}
					\label{fig:bipartite}
				\end{subfigure}
			} \vspace{-0.5cm}\caption{\footnotesize Different BP decoder factor 
				graph representations for a 
				$\mathcal{P}(8,4)$-code with information set $\mathbb{A}=\{4,6,7,8\}$.}
			\vspace{-0.7cm} \label{fig:bip}
		\end{figure}
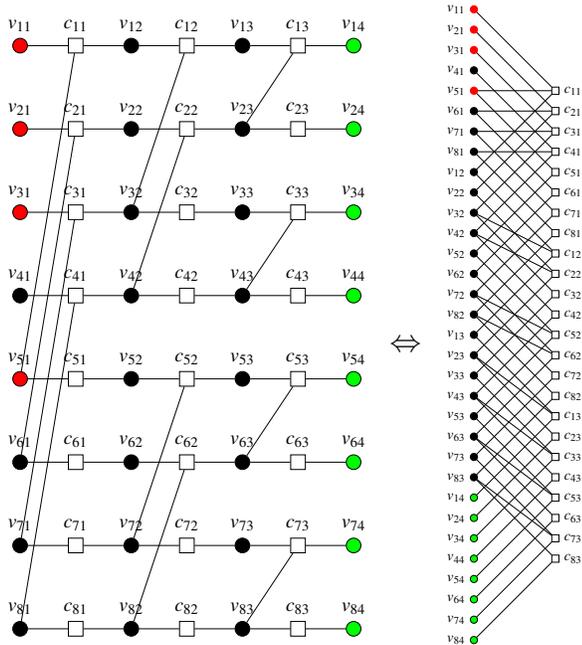	
		
		In this work, we show that one reason of the performance degradation when compared to the conventional \ac{BP} decoder is attributed to the 
		fact that the design of both -- polar code and its corresponding \ac{LDPC} 
		parity-check matrix -- do not take into consideration the new decoding schedule 
		(e.g., iterative manner, graph structure, scheduling, maximum number of 
		iterations $N_{it,max}$ etc.). In addition to that, in this paper, we optimize the pruned 
		parity-check matrix of the corresponding \ac{LDPC} code to outperform the 
		conventional \ac{BP} decoder of polar codes over a wide \ac{SNR} range and 
		approach the performance of the \ac{SCL} decoder at higher \ac{SNR} range. We 
		compare the newly obtained codes to the original ones to provide some reasoning 
		why the designed codes are better. We discuss this in terms of the weight 
		spectrum of the newly obtained codes in addition to the girth profile and the 
		degree distribution of the underlying parity-check matrices, which are also 
		shown to benefit from a reduced size and, thus, reduced complexity. Finally, 
		the special types of ``hidden'' nodes introduced in \cite{sparseBP} are interpreted as 
		punctured variable nodes and put into context within the design process.

		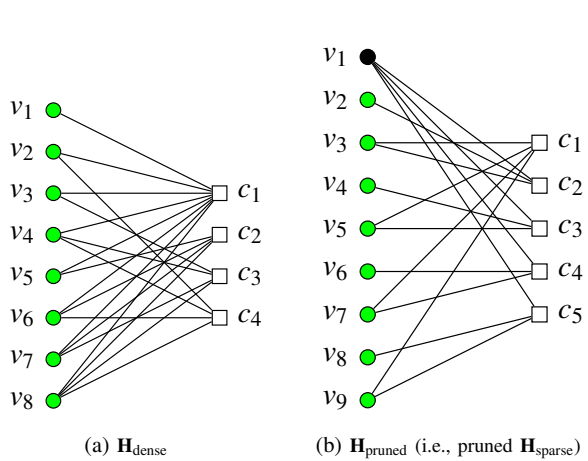
\begin{figure}[t]
			\resizebox{0.98\columnwidth}{!}{
				\begin{subfigure}[t]{0.5\columnwidth}\centering
					\resizebox{1\columnwidth}{!}{\tikzset{text=black, font={\fontsize{10pt}{10}\selectfont}}

\let\pgfmathMod=\pgfmathmod\relax


\def\n{3}  \def\N{8}
\begin{tikzpicture}
\tikzset{edge/.style = {-, thick}}

\tikzset{h1/.style={preaction={
draw,yellow,-,
double=yellow,
double distance=4\pgflinewidth,
}}}

\tikzset{h2/.style={preaction={
draw,green,-,
double=green,
double distance=4\pgflinewidth,
}}}
          \pgfmathtruncatemacro{\nn}{(\n + 1)}
%
%
%
%
%
%
%
%
%
%
%
%


\node [cnds, label=\cndPos:{$c_{1}$}](c1) at(12,3) {};
\node [cnds, label=\cndPos:{$c_{2}$}](c2) at(12,2.5) {};
\node [cnds, label=\cndPos:{$c_{3}$}](c3) at(12,2) {};
\node [cnds, label=\cndPos:{$c_{4}$}](c4) at(12,1.5) {};

\node [vndGs, label=\vndPos:{$v_{1}$}](v1) at(10,4) {};
\node [vndGs, label=\vndPos:{$v_{2}$}](v2) at(10,3.5) {};
\node [vndGs, label=\vndPos:{$v_{3}$}](v3) at(10,3) {};
\node [vndGs, label=\vndPos:{$v_{4}$}](v4) at(10,2.5) {};
\node [vndGs, label=\vndPos:{$v_{5}$}](v5) at(10,2) {};
\node [vndGs, label=\vndPos:{$v_{6}$}](v6) at(10,1.5) {};
\node [vndGs, label=\vndPos:{$v_{7}$}](v7) at(10,1) {};
\node [vndGs, label=\vndPos:{$v_{8}$}](v8) at(10,0.5) {};

\draw (c1)--(v1);
\draw (c1)--(v2);
\draw (c1)--(v3);
\draw (c1)--(v4);
\draw (c1)--(v5);
\draw (c1)--(v6);
\draw (c1)--(v7);
\draw (c1)--(v8);

\draw (c2)--(v5);
\draw (c2)--(v6);
\draw (c2)--(v7);
\draw (c2)--(v8);

\draw (c3)--(v3);
\draw (c3)--(v4);
\draw (c3)--(v7);
\draw (c3)--(v8);

\draw (c4)--(v2);
\draw (c4)--(v4);
\draw (c4)--(v6);
\draw (c4)--(v8);

\end{tikzpicture}
} 
					\caption{ \footnotesize $\mathbf{H}_{\text{dense}}$}
					\label{fig:dense}
				\end{subfigure}   

				\begin{subfigure}[t]{0.5\columnwidth}\centering
					
					\resizebox{\columnwidth}{!}{\tikzset{text=black, font={\fontsize{10pt}{10}\selectfont}}
\let\pgfmathMod=\pgfmathmod\relax
\definecolor{lightgrey}{gray}{0.9}

\begin{tikzpicture}[rotate=\angleRot]
\tikzset{edge/.style = {-, thick}}

\tikzset{h1/.style={preaction={draw,yellow,-,double=yellow,double distance=4\pgflinewidth,}}}

\tikzset{h2/.style={preaction={draw,green,-,double=green,double distance=4\pgflinewidth,}}}

\node [cnds, label=\cndPos:{$c_{1}$}](c1) at(3,2.5) {};

\node [cnds, label=\cndPos:{$c_{2}$}](c2) at(3,2) {};

\node [cnds, label=\cndPos:{$c_{3}$}](c3) at(3,1.5) {};

\node [cnds, label=\cndPos:{$c_{4}$}](c4) at(3,1) {};

\node [cnds, label=\cndPos:{$c_{5}$}](c5) at(3,0.5) {};

\node [vnds, label=\vndPos:{$v_{1}$}](v1) at(1,3.5) {};

\node [vndGs, label=\vndPos:{$v_{2}$}](v2) at(1,3) {};
\node [vndGs, label=\vndPos:{$v_{3}$}](v3) at(1,2.5) {};
\node [vndGs, label=\vndPos:{$v_{4}$}](v4) at(1,2) {};
\node [vndGs, label=\vndPos:{$v_{5}$}](v5) at(1,1.5) {};
\node [vndGs, label=\vndPos:{$v_{6}$}](v6) at(1,1) {};
\node [vndGs, label=\vndPos:{$v_{7}$}](v7) at(1,0.5) {};
\node [vndGs, label=\vndPos:{$v_{8}$}](v8) at(1,0) {};
\node [vndGs, label=\vndPos:{$v_{9}$}](v9) at(1,-0.5) {};

\draw (c1)--(v3);
\draw (c1)--(v7);

\draw (c2)--(v2);

\draw (c3)--(v4);

\draw (c4)--(v6);

\draw (c5)--(v8);

\draw (v1)--(c3);
\draw (v1)--(c5);

\draw (c2)--(v1);
\draw (c1)--(v5);
\draw (c4)--(v1);
\draw (c1)--(v9);
\draw (c2)--(v3);
\draw (c3)--(v5);
\draw (c4)--(v7);
\draw (c5)--(v9);

\end{tikzpicture}}
					\caption{ \footnotesize $\mathbf{H}_{\text{pruned}}$ (i.e., pruned
						$\mathbf{H}_{\text{sparse}}$)}
					\label{fig:pruning4}
				\end{subfigure}
				
			}
			\vspace{-0.2cm}			
			\caption{\footnotesize Dense vs. sparse/pruned Tanner graphs for a 
				$\mathcal{P}(8,4)$-code with information set $\mathbb{A}=\{4,6,7,8\}$.} 
			\label{fig:bipartite-factorgraph}
			\vspace{-0.7cm}
		\end{figure}
		Given the polarization matrix $\mathbf{G}_{N}$ of a polar code recursively constructed based on the $\mathbf{G}_2$ polarizing kernel, 
		the parity-check matrix denoted as $\mathbf{H}_{\text{dense}}$ is formed from the columns of $\mathbf{G}_N$ with indices in $\bar{\mathbb{A}}$, where $\bar{\mathbb{A}}$ is the set of frozen indices, as proven in \cite[Lemma 1]{LPpolar}. Note that, the corresponding factor graph of $\mathbf{H}_{\text{dense}}$ is not sparse and, thus, it is not possible to apply the traditional decoding algorithms (e.g., SPA decoding) on the naive $\mathbf{H}_{\text{dense}}$, see Fig. \ref{fig:dense}.

		As depicted in \cite[Fig. 4a]{sparseBP}, the $\log_2(N)+1$ sets of variable 
		nodes (VNs) and $\log_2(N)$ sets of check nodes (CNs) in Arıkan's \ac{BP} 
		factor graph can be re-grouped into a bipartite graph consisting of \emph{only} 
		two sets: VNs $\mathcal{V}$ and CNs $\mathcal{C}$ \cite{LPpolar}. Assuming the 
		schedule is now the same as the conventional flooding BP of LDPC codes, the 
		graph resembles an \ac{LDPC} code with some special constraint. In 
		\cite{sparseBP}, it was given the name  ``LDPC-like'' to address the fact that 
		the channel information bits of this \ac{LDPC} code are only at the last $N$ 
		VNs. Applying some basic pruning, the graph can be significantly reduced 
		with an approximate reduction factor of $80\%$. However, the final compact 
		factor graph still contains a non-negligible portion of such nodes, called 
		``hidden VNs'', corresponding to the black VNs in Fig. \ref{fig:pruning4}.
		In this work, these nodes are interpreted as punctured parity VNs of 
		the \ac{LDPC} code. Thus, taking into consideration the connections of these 
		punctured nodes to the rest of the graph while designing the code has an impact 
		on the performance of the whole code as shown later. For more details on the 
		pruning of the originally large bipartite graph, we refer the interested reader 
		to \cite[Sec.~III]{sparseBP} and the detailed source code provided online 
		\cite{github-ldpc-like-polar}.

		\section{Code Construction Mismatch Paradigm} \label{sec:genAlg}

		Polar code construction is the process of choosing the best $k$ synthesized bit-channels out of the $N$ synthesized bit-channels. The best $k$ synthesized bit-channels are included in the information set $\mathbb{A}$ and are used for data transmission. Most of the state-of-the-art polar code construction algorithms are tailored to the hard-output 
		\ac{SC} decoder. For the case of an \ac{AWGN} channel, they are based 
		either on bounds (e.g., the Bhattacharyya 
		parameter \cite{ArikanMain}, density evolution 
		\cite{constructDE}), approximations (e.g., Gaussian approximation (GA)
		\cite{constructGaussian}), or heuristics (e.g., polarization 
		weight (PW) \cite{PW} and $\beta$-expansion \cite{BetaIngmard}). It is worth-mentioning that an efficient density evolution-based implementation was proposed in \cite{constructTalVardy}. There also 
		exist several Monte-Carlo-based designs for specific decoding schemes. 
		Additionally, various schemes focus on improving the finite-length performance 
		of polar codes either by \emph{interpolating} them using other codes (e.g., \ac{RM}-codes), to enjoy good finite-performance provided by these codes \cite{HybridTse,RMurbankePolar} or via 
		concatenation 
		schemes with other codes, e.g., \cite{subCodes, BP_Siegel_Concatenating, 
			BP_felxible}.
		
		Furthermore, both iterative-specific parameters  of the family of iterative 
		decoders (e.g., 
		scheduling, stopping sets, cycles, girths, etc.) and the  list decoding fashion of 
		\ac{SCL} decoder variants are not usually considered in the polar code 
		construction phase \cite{polarDesign5G}. 
		Therefore, this leaves the door wide open for optimizing the information set $\mathbb{A}$ and tailoring it to the decoder, thus, utilizing the decoder at the most efficient manner. This remark is even more significant in the finite-length regime where polar codes suffer from a degraded performance due to semi-polarized bit channels. 
		

		In \cite{GenAlg_Journal_IEEE}, a new code construction framework for polar codes
		was presented where the Genetic Algorithm (GenAlg) was applied to the code 
		construction optimization problem on a specific error-rate simulation 
		environment, i.e., taking into consideration the actual decoder and channel. 
		This \emph{decoder-in-the-loop} design was shown to yield significant 
		gains in terms of error-rate performance, where \ac{SCL} performance was 
		enhanced to achieve that of the \ac{CA-SCL} even \emph{without} using a \ac{CRC} code, as 
		shown in \cite[Fig. 6]{GenAlg_Journal_IEEE}.
		
		Inspired by that, the GenAlg is applied to tailor the LDPC-like code, derived from the polar code, to the LDPC-style decoder and take into account its special constraints (i.e., the big set of punctured variable nodes). Due to limited space, we refer the interested reader to \cite{GenAlg_Journal_IEEE} and the pseudo-algorithms therein for specific details about the GenAlg framework.
		
			\vspace{-0.2cm}		
		\section{Results and Discussion}
					\vspace{-0.1cm}
		In this section, numerical results are presented in terms of both BER and BLER, as cost functions, to demonstrate the flexibility of our proposed algorithm. The optimized codes are compared at different lengths with standardized codes of the same corresponding block length and code rate. 
		All considered polar and LDPC codes are simulated over the binary-input {AWGN} channel.
        In the following, by 5G polar codes we mean the bit-reliability order of the nested polar code with maximum code length $N_{max}=1024$, specified by the 3GPP group in \cite{polar5G2018}.		
		
					\vspace{-0.15cm}
		
		\subsection{Codeword length $N=128$}
		We design a $\mathcal{P}(128,64)$ polar code and its corresponding pruned 
		sparse $\mathbf{H}_{\text{pruned}}$ under LDPC-like decoding. All reference 
		LDPC codes from \cite{liva2016} are designed through a girth optimization 
		technique based on the PEG algorithm \cite{PEG}: the standardized LDPC code by 
		the Consultative Committee for Space Data Systems (CCSDS) for satellite telecommand 
		links, an accumulate-repeat-3-accumulate (AR3A) LDPC code and an 
		accumulate-repeat-jagged-accumulate (ARJA) LDPC code. A further reference code 
		is the 5G polar code without the CRC-aid. As 
		depicted in  Fig. \ref{fig:BER128}, the BLER performance of the BLER-optimized polar code under 
		LDPC(-like) decoding achieves the performance of the 5G polar code  under 
		Arıkan's conventional BP decoder, with a performance gain of $0.25$ dB 
		at BLER of $10^{-2}$ compared to the originally proposed code in 
		\cite{sparseBP}. It outperforms  the 5G polar code  under Arıkan's 
		conventional BP decoder in the higher SNR range. 
		\begin{figure}[t]
			\centering
			\resizebox{\columnwidth}{!}{
%
%
\definecolor{mycolor1}{rgb}{0.46600,0.67400,0.18800}%
\definecolor{mycolor2}{rgb}{0.92900,0.69400,0.12500}%
\definecolor{mycolor3}{rgb}{0.49400,0.18400,0.55600}%
\definecolor{mycolor4}{rgb}{0.00000,0.49804,0.00000}%
\definecolor{mycolor5}{rgb}{0.50000,0.44700,0.74100}%

\begin{tikzpicture}

\begin{axis}[%
width=\columnwidth,
height=0.75\columnwidth,
at={(2.516in,1.149in)},
scale only axis,
xmin=1,
xmax=3,
xtick={1,1.5,2,2.5,3},
xlabel style={font=\color{white!15!black}},
xlabel={\large $E_b/N_0$ [dB]},
ymode=log,
ymin=0.01,
ymax=1,
yminorticks=true,
ylabel style={    font={\color{white!15!black}}},
ylabel={BLER},
axis background/.style={fill=white},
xmajorgrids,
xminorgrids,
ymajorgrids,
yminorgrids,
mark options={solid},
legend style={at={(0.5,1.03)}, anchor=south, legend cell align=left, align=left, draw=white!15!black}
]

\addplot [color=mycolor3, mark=x, dashed, line width=1.5pt,  mark options={solid, mycolor3}]
  table[row sep=crcr]{%
1	0.6061\\
1.5	0.3333\\
2	0.1869\\
2.5	0.07519\\
3	0.02117\\
3.5	0.004001\\
4	0.0007755\\
4.5	0.0001107\\
5	1.8e-05\\
5.75	8e-07\\
};
\label{plot128bler:regLDPC}

\addplot [color=blue, dashed, line width=1.5pt, mark=diamond, mark options={solid, blue}]
table[row sep=crcr]{%
	1	0.5089\\
	1.5	0.2532\\
	2	0.1241\\
	2.5	0.04344\\
	3	0.01137\\
	3.5	0.002291\\
	4	0.0003477\\
	4.5	5.222e-05\\
	5	5e-06\\
	5.4	1.4e-06\\
};
\label{plot128bler:ar3aLDPC}

\addplot [color=mycolor2, dashed, line width=1.5pt, mark=o, mark options={solid, mycolor2}]
table[row sep=crcr]{%
	1	0.625\\
	1.5	0.3683\\
	2	0.1828\\
	2.5	0.07855\\
	3	0.02052\\
	3.5	0.004433\\
	4	0.0007295\\
	4.5	7.811e-05\\
	5	8.671e-06\\
	5.5	7e-07\\
};
\label{plot128bler:arjaLDPC}

\addplot [color=mycolor5, dashed, line width=1.5pt, mark=star, mark options={solid, mycolor5}]
table[row sep=crcr]{%
	1	0.76965\\
	1.5	0.56176\\
	2	0.33481\\
	2.5	0.15143\\
	3	0.04967\\
	3.5	0.0115768\\
	4	0.0018724\\
	4.5	0.000209666666666667\\
	5	1.684e-05\\
	5.5	1.05e-06\\
	6	8e-08\\
};
\label{plot128bler:ccsdsLDPC}

\addplot [color=red, line width=1.5pt]
table[row sep=crcr]{%
	1	0.426790423768448\\
	1.3887151667869	0.27201316949068\\
	1.79564462090408	0.149061386814392\\
	2.22257959726683	0.070392286745732\\
	2.6715892883487	0.0292727362569996\\
	3.14508150615463	0.0109446451569739\\
	3.64588085277352	0.0037044093820294\\
	4.17733083711007	0.00107631634230462\\
	4.74342931819689	0.000250240536353706\\
	5.34901124377515	4.26210573008244e-05\\
	6	4.86465415083642e-06\\
};
\label{plot128bler:5GBP}

\addplot [color=black, line width=1.5pt, mark=square, mark options={solid, black}]
  table[row sep=crcr]{%
1	0.45626\\
1.5	0.26589\\
2	0.12835\\
2.5	0.05236\\
3	0.01841\\
};

\label{plot128bler:5Gldpclike}

\addplot [color=mycolor4, line width=1.5pt, mark=o, mark options={solid, mycolor4}]
  table[row sep=crcr]{%
1	0.438\\
1.5	0.244\\
2	0.113\\
2.5	0.041\\
3	0.011\\
};
\label{plot128bler:genAlgldpclike}

 \coordinate (legend) at (axis description cs:0.5,1.05);
\end{axis}

 \matrix [
draw,
matrix of nodes,
anchor=south,
mark options={solid}
] at (legend) {
							         & Code 					                   & Decoder      \\
	\ref{plot128bler:regLDPC}        & Regular (3,6) LDPC     			  		   &  LDPC   	    \\
	\ref{plot128bler:ar3aLDPC}       & AR3A LDPC  from \cite{liva2016}             &  LDPC  	    \\
	\ref{plot128bler:arjaLDPC}       & ARJA LDPC  from \cite{liva2016}      &  LDPC   	    \\
	\ref{plot128bler:ccsdsLDPC} 	 & CCSDS LDPC                                  &  LDPC          \\
	\ref{plot128bler:5GBP}  		 & 5G polar                                    &  Arıkan's BP   \\
	\ref{plot128bler:5Gldpclike}     & 5G polar                                    &  LDPC(-like)   \\
	\ref{plot128bler:genAlgldpclike} & Optimized                                   &  LDPC(-like)   \\	
};

\end{tikzpicture}%
} 
			\vspace{-0.4cm}
			%
			%
			%
			%
			\vspace{-0.35cm}
			\caption{\footnotesize BLER performance of the \emph{optimized} 
				$\mathcal{P}(128,64)$-code over the \ac{AWGN} channel at 
				$\mathrm{SNR_{des}}=\unit[3]{dB}$ under 
				LDPC(-like) decoding. All iterative decoders use $N_{it,max}=200$.}\label{fig:BER128}	
			\vspace{-0.6cm}
		\end{figure}
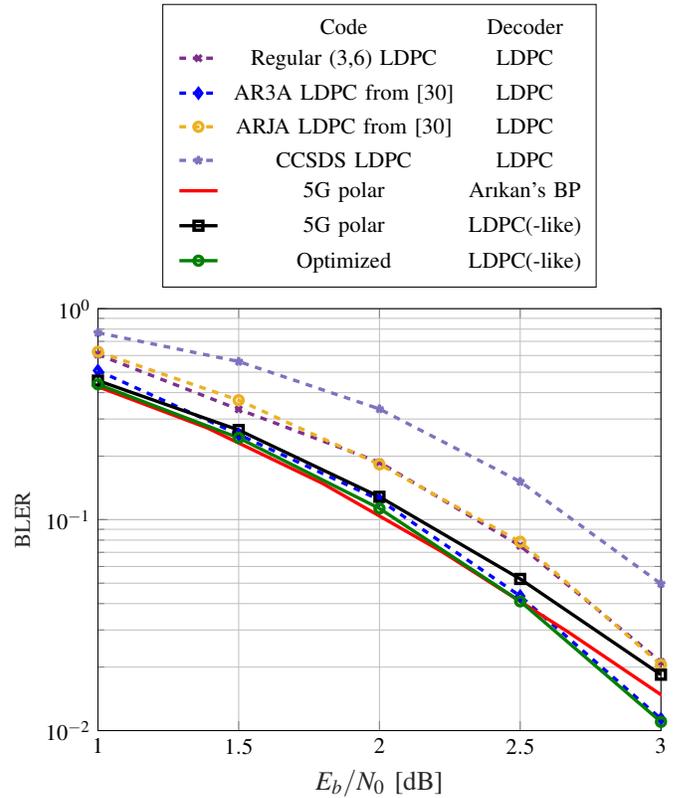

		Besides, it outperforms the LDPC codes presented in Fig.~\ref{fig:BER128}. It is, however, important to keep in mind that these codes have special constraints on their structure for implementation reasons (i.e., encoding complexity issues), yet, the message here is to show that the newly proposed codes are of comparable performance  which brings them to attention for further improvements. Note that, the proposed code is encoded via a low-complexity polar encoder, despite being decoded by a conventional LDPC decoder.
		
		Tab. \ref{tab:A_dmin1} depicts a comparison between the minimum distance 
		of both codes. It turns out that the optimized code has 
		the same minimum distance as the initial 5G code, however, at a significantly reduced 
		number of minimum-weight codewords $A_{d_{min}}$. This was computed using 
		the algorithm 
		proposed in \cite{AdaptiveList} with a list size of up to $L=5\cdot 10^5$.
		

		One more comparison providing insights on why the obtained code performs better is 
		the girth of the optimized $\mathbf{H}$-matrix. Both $\mathbf{H}$-matrices of the 
		initial 5G and the optimized code are of minimum girth-6. However, the 
		optimized code has a reduced number of 1825 cycles of girth-6, when compared to 
		the original $\mathbf{H}$-matrix corresponding to the 5G polar code of 2234 
		girth-6 cycles. Furthermore, the optimized $\mathbf{H}_{\text{opt}}$-matrix 
		has a reduced 
		size of $365\times493$ compared to an original size of 
		$\mathbf{H}_{\text{pruned,5G}}$ 
		of $503\times 633$. Therefore, the number of punctured variable nodes has been minimized.
			\vspace{-0.2cm}
		\begin{table}[H]
			\fontsize{9}{12}\selectfont 
			\begin{center}
				\caption{\footnotesize{The number of minimum-weight codewords of a 
						$\mathcal{P}$(128,64)-code}}	\vspace{-0.1cm}
				\label{tab:A_dmin1}
				\begin{tabular}{ccc}
					\hline 
					Construction  & $d_{min}$ & $A_8$\tabularnewline
					\hline 
					$5$G \cite{polar5G2018} & 8 & 304 \tabularnewline
					
					Optimized @ $\unit[3]{dB}$ & 8 & 170\tabularnewline
					\hline 
					\vspace{-0.8cm}
				\end{tabular}
			\end{center}
		\end{table}
		
			\vspace{-0.2cm}		

		\subsection{Codeword length $N=256$}
		\begin{figure}[t]
			\vspace{-0.4cm}
			\centering
			\resizebox{\columnwidth}{!}{
%
%
\definecolor{mycolor1}{rgb}{0.50196,0.50196,0.50196}%
\definecolor{mycolor2}{rgb}{0.00000,0.49804,0.00000}%
\definecolor{mycolor3}{rgb}{0.07451,0.12157,0.12157}%
\definecolor{mycolor4}{rgb}{0.65098,0.65098,0.65098}%
\definecolor{mycolor5}{rgb}{0.92900,0.69400,0.12500}%
\begin{tikzpicture}

\begin{axis}[%
width=\columnwidth,
height=0.8\columnwidth,
at={(2.454in,1.062in)},
scale only axis,
xmin=1,
xmax=3,
xtick={1,1.5,2,2.5,3},
xlabel style={font=\color{mycolor3}},
xlabel={\large $E_b/N_0$ [dB]},
ymode=log,
ymin=0.0005,
ymax=1,
yminorticks=true,
ylabel style={font=\color{mycolor3}},
ylabel={BER / BLER},
axis background/.style={fill=white},
xmajorgrids,
xminorgrids,
ymajorgrids,
yminorgrids,
legend style={at={(0.5,1.03)}, anchor=south, legend cell align=left, align=left, draw=white!15!black}
]

\addplot [color=red, dashed, line width=1.5pt, mark=square, mark options={solid, red}]
table[row sep=crcr]{%
	1	0.472851095159212\\
	1.43299946165632	0.241286645612269\\
	1.88872205947323	0.0960108474576271\\
	2.36968547635921	0.0301409372353646\\
	2.87885049445184	0.00824433042461644\\
	3.41973148833631	0.00206549566245911\\
	3.99654366295697	0.00046447427205765\\
	4.61440275098179	7.73998839001741e-05\\
	5.27960133120302	9.70438444089039e-06\\
	6	1.00040556441581e-06\\
};
\label{plot256bler:5GBP}

\addplot [color=blue, dashed, line width=1.5pt, mark=o, mark options={solid, blue}]
  table[row sep=crcr]{%
1	0.5256\\
1.5	0.27047\\
2	0.10545\\
2.5	0.03148\\
3	0.00831\\
};
\label{plot256bler:5Gldpclike}

\addplot [color=mycolor2, dashed, line width=1.5pt, mark=diamond*, mark options={solid, fill=mycolor2, mycolor2}]
table[row sep=crcr]{%
	1	0.49662\\
	1.5	0.24187\\
	2	0.08305\\
	2.5	0.01982\\
	3	0.00412\\
};
\label{plot256bler:genAlgldpclike}

\addplot [color=mycolor1, line width=1.5pt]
table[row sep=crcr]{%
	6.72754546996622	0\\
	6.15085245675472	1.13990547412772e-07\\
	5.61007623760508	8.36088562632378e-07\\
	5.10100406465991	3.13671028088224e-06\\
	4.62012349103394	1.40757601363598e-05\\
	4.16446123883273	4.58412412215316e-05\\
	3.73152960129824	0.000130557446939149\\
	3.31915599482783	0.000344328114295337\\
	2.92547618737117	0.000853919215127727\\
	2.54886856308915	0.00207535486962485\\
	2.18791376729822	0.00493247301690723\\
	1.84136240797438	0.0112997255291268\\
	1.50810895758506	0.0240736063435768\\
	1.18717047957507	0.0461151032889553\\
	0.877669147263633	0.0813686407083195\\
	0.578817771721302	0.126336477510693\\
	0.289907737741013	0.179743577023949\\
	0.0102988825148742	0.234530602730165\\
};
\label{plot256ber:5Gscl}

\addplot [color=red, line width=1.5pt, mark=square, mark options={solid, red}]
table[row sep=crcr]{%
	1	0.143811976578318\\
	1.43299946165632	0.0634307919552175\\
	1.88872205947323	0.0211869279661017\\
	2.36968547635921	0.00537011103343891\\
	2.87885049445184	0.00116830570792511\\
	3.41973148833631	0.000235157318669631\\
	3.99654366295697	4.3653120917902e-05\\
	4.61440275098179	6.74139613790579e-06\\
	5.27960133120302	8.46476187941712e-07\\
	6	7.19041499423866e-08\\
};
\label{plot256ber:5GBP}

\addplot [color=blue, line width=1.5pt, mark=o, mark options={solid, blue}]
table[row sep=crcr]{%
	1	0.1739796875\\
	2	0.02424921875\\
	3	0.00185390625\\
};
\label{plot256ber:5Gldpclike}

\addplot [color=mycolor2, line width=1.5pt, mark=diamond, mark options={solid, mycolor2}]
table[row sep=crcr]{%
	1	0.1316\\
	1.5	0.0561\\
	2	0.01695\\
	2.5	0.0036\\
	3	0.00060133\\
};
\label{plot256ber:genAlgldpclike}

 \coordinate (legend) at (axis description cs:0.5,1.05);
\end{axis}

\matrix [
draw,
matrix of nodes,
anchor=south,
mark options={solid}
] at (legend) {
                                   	 & Code	     	   & Decoder         &	           \\
	\ref{plot256bler:5GBP}           & 5G polar        &  Arıkan's BP    & BLER  	    \\
	\ref{plot256bler:5Gldpclike}     & 5G polar        &  LDPC(-like)    & BLER 	    \\
	\ref{plot256bler:genAlgldpclike} & Optimized       &  LDPC(-like)    & BLER  	    \\
	\ref{plot256ber:5Gscl} 	         & 5G polar        &  SCL ($L=128$)  & BER         \\
	\ref{plot256ber:5GBP}  		     & 5G polar        &  Arıkan's BP    & BER 			\\
	\ref{plot256ber:5Gldpclike}      & 5G polar        &  LDPC(-like)    & BER  		\\
	\ref{plot256ber:genAlgldpclike}  & Optimized       &  LDPC(-like)    & BER  		\\	
};

\end{tikzpicture}%
} 
			\vspace{-0.25cm}
			%
			%
			%
			\vspace{-0.45cm}
			\caption{\footnotesize BER/BLER performance of the \emph{optimized} 
				$\mathcal{P}(256,128)$-codes  over the \ac{AWGN} channel at 
				$\mathrm{SNR_{des}}=\unit[3]{dB}$ under 
				LDPC(-like) decoding. All iterative decoders 
				use $N_{it,max}=200$ and no CRC is used.}\label{fig:BER256}	
			\vspace{-0.55cm}
		\end{figure}
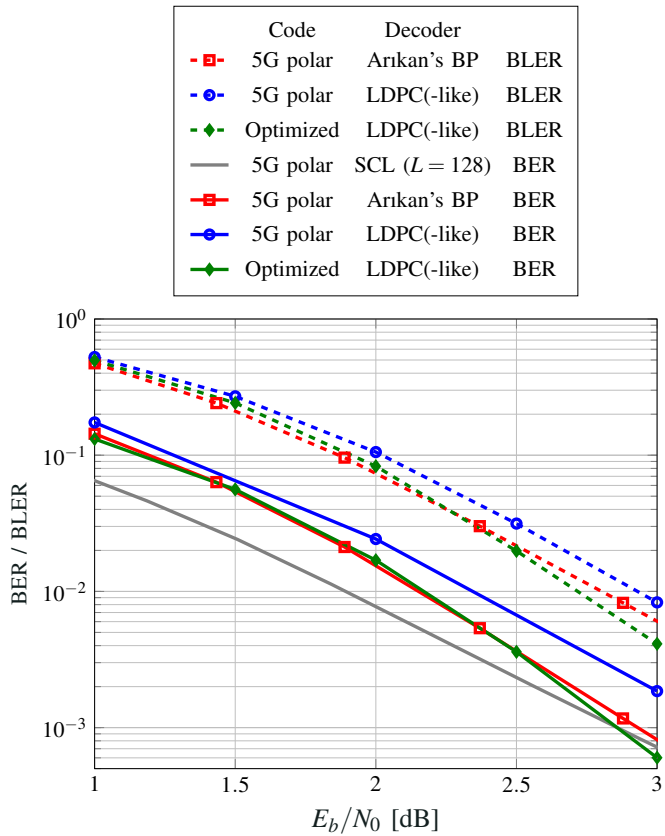
		We design a $\mathcal{P}(256,128)$ polar code and its corresponding pruned sparse $\mathbf{H}_{\text{pruned}}$ under LDPC-like decoding with both BLER and BER as cost functions. The newly obtained codes are compared to the 5G polar code without the CRC-aid \cite{polar5G2018} under BP, LDPC-like and plain  \ac{SCL} decoding. As depicted in  Fig. \ref{fig:BER256}, both BER and BLER performance of the optimized codes under LDPC(-like) decoding achieve the performance of the 5G polar code  under Arıkan's conventional BP decoder, with a performance gain of around $0.2$ dB and $0.1$ dB at BLER of $10^{-2}$ and BER of $10^{-3}$, respectively, compared to the originally proposed code in \cite{sparseBP}. They even outperform  the 5G polar code  under both Arıkan's conventional BP \emph{and} plain \ac{SCL} decoders towards a higher SNR range. 
		
		Tab. \ref{tab:A_dmin} depicts a comparison between the minimum distance 
		of both codes (initial and BLER-optimized). It turns out that the 
		BLER-optimized code has the same minimum distance as the initial 5G polar code, 
		with significantly reduced number of minimum-weight codewords. These numbers were obtained using the algorithm in \cite{AdaptiveList} with a list size 
		up to $L=2.5\cdot 10^5$.
		
		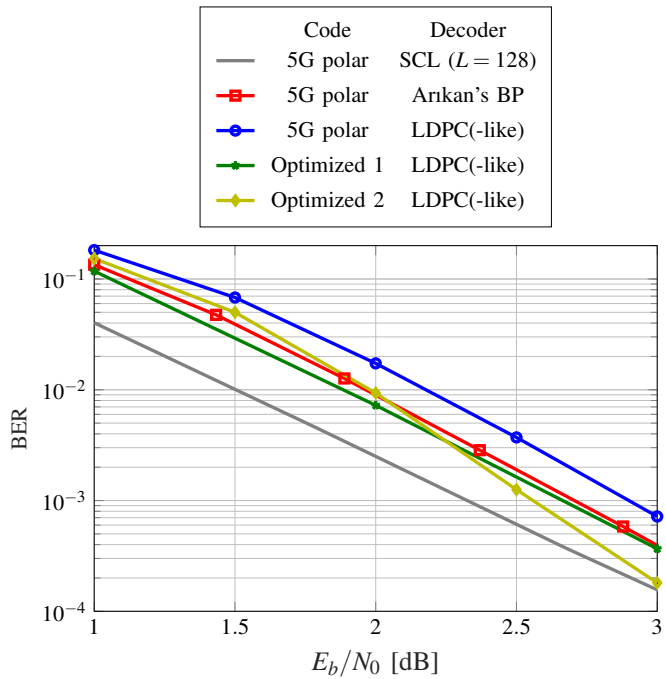
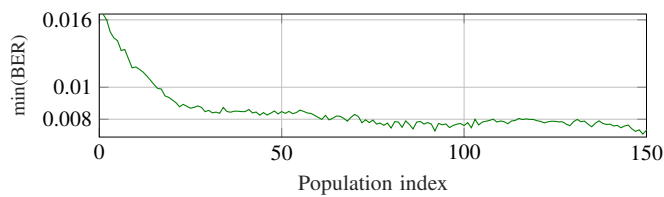
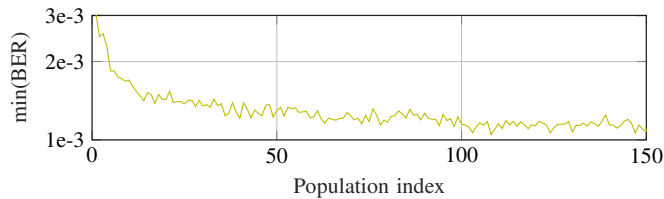
\begin{figure}[t]
			\vspace{-0.3cm}
			\centering
			\begin{subfigure}[t]{\columnwidth}
				\resizebox{\columnwidth}{!}{
%
%
\definecolor{mycolor1}{rgb}{0.50196,0.50196,0.50196}%
\definecolor{mycolor2}{rgb}{0.00000,0.49804,0.00000}%
\definecolor{mycolor3}{rgb}{0.74902,0.74902,0.00000}%
\begin{tikzpicture}

\begin{axis}[%
width=\columnwidth,
height=0.65\columnwidth,
at={(2.516in,1.106in)},
scale only axis,
xmin=1,
xmax=3,
xtick={1,1.5,2,2.5,3},
xtick={1,1.5,2,2.5,3},
xlabel style={font=\color{white!15!black}},
xlabel={\large $E_b/N_0$ [dB]},
ymode=log,
ymin=0.0001,
ymax=0.2,
yminorticks=true,
ylabel style={font=\color{white!15!black}},
ylabel={BER},
axis background/.style={fill=white},
xmajorgrids,
ymajorgrids,
yminorgrids,
legend style={at={(0.5,1.03)}, anchor=south, legend cell align=left, align=left, draw=white!15!black}
]

\addplot [color=mycolor1, line width=1.5pt]
table[row sep=crcr]{%
	1	0.0402415478765871\\
	1.79564462090408	0.00445505603741749\\
	2.6715892883487	0.000375954273950299\\
	3.64588085277352	2.7788150645692e-05\\
	4.74342931819689	1.47239947385341e-06\\
	6	1.68041511974685e-08\\
};
\label{plot512ber:5Gscl}

\addplot [color=red, line width=1.5pt, mark=square, mark options={solid, red}]
  table[row sep=crcr]{%
1	0.134603902131215\\
1.43299946165632	0.0474489977954841\\
1.88872205947323	0.0126516415227093\\
2.36968547635921	0.00284784485052425\\
2.87885049445184	0.000583087589044871\\
3.41973148833631	0.000101105418197291\\
3.99654366295697	1.69208598562669e-05\\
4.61440275098179	2.94218249828975e-06\\
5.27960133120302	2.85546789335963e-07\\
6	3.75152612082595e-08\\
};
\label{plot512ber:5GBP}

\addplot [color=blue, line width=1.5pt, mark=o, mark options={solid, blue}]
  table[row sep=crcr]{%
1	0.1822421875\\
1.5	0.06803828125\\
2	0.01730703125\\
2.5	0.0037140625\\
3	0.00071796875\\
};
\label{plot512ber:5Gldpclike}

\addplot [color=mycolor2, line width=1.5pt, mark=star, mark options={solid, mycolor2}]
  table[row sep=crcr]{%
1	0.11777265625\\
2	0.0072390625\\
3	0.0003671875\\
};
\label{plot512ber:genAlg1ldpclike}

\addplot [color=mycolor3, line width=1.5pt, mark=diamond, mark options={solid, mycolor3}]
  table[row sep=crcr]{%
1	0.1530178515625\\
1.5	0.0499884375\\
2	0.0093421484375\\
2.5	0.0012555859375\\
3	0.0001807421875\\
};

\label{plot512ber:genAlg2ldpclike}

 \coordinate (legend) at (axis description cs:0.5,1.05);
\end{axis}

\matrix [
draw,
matrix of nodes,
anchor=south,
mark options={solid}
] at (legend) {
	& Code 										    	& Decoder               \\
	\ref{plot512ber:5Gscl} 	          & 5G polar    &  SCL ($L=128$)        \\
	\ref{plot512ber:5GBP}  		      & 5G polar    &  Arıkan's BP     		\\
	\ref{plot512ber:5Gldpclike}       & 5G polar    &  LDPC(-like)   	\\
	\ref{plot512ber:genAlg1ldpclike}  & Optimized  1&  LDPC(-like)   	\\	
	\ref{plot512ber:genAlg2ldpclike}  & Optimized  2&  LDPC(-like)   	\\	
};

\end{tikzpicture}%
} 
				\vspace{-0.6cm}\caption{ \footnotesize BER comparison.  All iterative decoders use 
					$N_{it,max}=200$.} \label{fig:BER512}\vspace{0.25cm}
			\end{subfigure}
			
			\begin{subfigure}{\columnwidth}
				
				\resizebox{\columnwidth}{!}{
%
%
\definecolor{mycolor1}{rgb}{0.00000,0.49804,0.00000}%
%
\begin{tikzpicture}

\begin{axis}[%
width=\columnwidth,
height=0.225\columnwidth,
at={(2.516in,1.149in)},
scale only axis,
xmin=0,
xmax=150,
xlabel style={font=\color{white!15!black}},
xlabel={Population index},
ymode=log,
ymin=0.0070646875,
ymax=0.016662578125,
ytick={0.007, 0.008, 0.01, 0.016},
yticklabels={0.007, 0.008, 0.01, 0.016},
xtick={0,50,100,150},
yminorticks=true,
ylabel style={font=\color{white!15!black}},
ylabel={\small min(BER)},
axis background/.style={fill=white},
xmajorgrids,
xminorgrids,
ymajorgrids,
yminorgrids,
legend style={legend cell align=left, align=left, draw=white!15!black}
]
\addplot [color=mycolor1, line width=0.5pt]
  table[row sep=crcr]{%
1	0.016662578125\\
2	0.016016640625\\
3	0.014711484375\\
4	0.014099921875\\
5	0.01383625\\
6	0.01292734375\\
7	0.013005234375\\
8	0.012190234375\\
9	0.011450390625\\
10	0.0115203125\\
11	0.011333203125\\
12	0.011115234375\\
13	0.0108309375\\
14	0.0105084375\\
15	0.010193046875\\
16	0.009914609375\\
17	0.009878203125\\
18	0.009416875\\
19	0.00932078125\\
20	0.009151015625\\
21	0.008984921875\\
22	0.008725625\\
23	0.00887234375\\
24	0.00875609375\\
25	0.00863953125\\
26	0.00869015625\\
27	0.00878015625\\
28	0.008692578125\\
29	0.00843375\\
30	0.008509609375\\
31	0.008363203125\\
32	0.008407109375\\
33	0.00833953125\\
34	0.00868359375\\
35	0.008442578125\\
36	0.00840109375\\
37	0.008464296875\\
38	0.00846109375\\
39	0.008436171875\\
40	0.00843796875\\
41	0.0085621875\\
42	0.00835984375\\
43	0.008410546875\\
44	0.008230625\\
45	0.00837953125\\
46	0.00823859375\\
47	0.008345\\
48	0.0084734375\\
49	0.00829890625\\
50	0.00843828125\\
51	0.008335390625\\
52	0.00845046875\\
53	0.008318046875\\
54	0.008376484375\\
55	0.008526875\\
56	0.008453203125\\
57	0.008358125\\
58	0.008332734375\\
59	0.008224375\\
60	0.0081015625\\
61	0.0079940625\\
62	0.00821875\\
63	0.007966640625\\
64	0.008057734375\\
65	0.008177109375\\
66	0.00816640625\\
67	0.008051875\\
68	0.007893984375\\
69	0.008110625\\
70	0.008271171875\\
71	0.008156328125\\
72	0.0078075\\
73	0.007967578125\\
74	0.00781640625\\
75	0.007955625\\
76	0.00773703125\\
77	0.00778578125\\
78	0.007668828125\\
79	0.007788828125\\
80	0.0075259375\\
81	0.0078715625\\
82	0.0078425\\
83	0.0075640625\\
84	0.007909609375\\
85	0.007717109375\\
86	0.00747828125\\
87	0.007840859375\\
88	0.007878828125\\
89	0.007736640625\\
90	0.0078184375\\
91	0.007741640625\\
92	0.007375390625\\
93	0.0077703125\\
94	0.007700078125\\
95	0.00774265625\\
96	0.007549375\\
97	0.00766\\
98	0.00773140625\\
99	0.00776515625\\
100	0.00765921875\\
101	0.007810078125\\
102	0.007539921875\\
103	0.007998359375\\
104	0.00768921875\\
105	0.007844140625\\
106	0.007890546875\\
107	0.00793765625\\
108	0.00800046875\\
109	0.00784703125\\
110	0.007904765625\\
111	0.007861796875\\
112	0.007751953125\\
113	0.007905859375\\
114	0.00793140625\\
115	0.00804359375\\
116	0.007981796875\\
117	0.008019140625\\
118	0.007990546875\\
119	0.007987265625\\
120	0.0079178125\\
121	0.00788\\
122	0.0078021875\\
123	0.007858984375\\
124	0.007891328125\\
125	0.007885\\
126	0.00785546875\\
127	0.007857890625\\
128	0.007734140625\\
129	0.0076475\\
130	0.007864453125\\
131	0.007987421875\\
132	0.0078675\\
133	0.007897421875\\
134	0.00772703125\\
135	0.00759\\
136	0.007786953125\\
137	0.00791140625\\
138	0.007771328125\\
139	0.00771890625\\
140	0.0077275\\
141	0.007633515625\\
142	0.00766765625\\
143	0.007539453125\\
144	0.00763296875\\
145	0.00768046875\\
146	0.00748078125\\
147	0.0073509375\\
148	0.007430546875\\
149	0.007214296875\\
150	0.0074140625\\
151	0.007477421875\\
152	0.007358203125\\
153	0.007325625\\
154	0.00755421875\\
155	0.00747015625\\
156	0.00746828125\\
157	0.007508125\\
158	0.007448984375\\
159	0.007421796875\\
160	0.0074228125\\
161	0.007260546875\\
162	0.00743109375\\
163	0.007366328125\\
164	0.0073940625\\
165	0.007475546875\\
166	0.007334453125\\
167	0.007461328125\\
168	0.00743765625\\
169	0.007444609375\\
170	0.00729625\\
171	0.00748828125\\
172	0.00736375\\
173	0.00745671875\\
174	0.00754265625\\
175	0.007346015625\\
176	0.0073146875\\
177	0.007298671875\\
178	0.00742125\\
179	0.00743625\\
180	0.0076565625\\
181	0.007368203125\\
182	0.00736625\\
183	0.00739796875\\
184	0.007383515625\\
185	0.00734140625\\
186	0.007490546875\\
187	0.007543203125\\
188	0.00755640625\\
189	0.007579609375\\
190	0.00770625\\
191	0.007556171875\\
192	0.007612578125\\
193	0.00753484375\\
194	0.007620625\\
195	0.007405546875\\
196	0.007453359375\\
197	0.00742625\\
198	0.007560546875\\
199	0.007286953125\\
200	0.0073334375\\
201	0.00731328125\\
202	0.00738578125\\
203	0.007469453125\\
204	0.0074378125\\
205	0.00747578125\\
206	0.007459296875\\
207	0.007311484375\\
208	0.00743546875\\
209	0.007496484375\\
210	0.0075940625\\
211	0.0075225\\
212	0.007539453125\\
213	0.007443828125\\
214	0.0073528125\\
215	0.007399453125\\
216	0.00729875\\
217	0.00752375\\
218	0.0075003125\\
219	0.00744859375\\
220	0.00725203125\\
221	0.007278359375\\
222	0.0075359375\\
223	0.00733296875\\
224	0.007270390625\\
225	0.0070646875\\
226	0.007462421875\\
227	0.007352421875\\
228	0.007224375\\
229	0.00745875\\
230	0.007373359375\\
231	0.00715015625\\
232	0.00717609375\\
233	0.007391484375\\
234	0.0072759375\\
235	0.00734875\\
236	0.007249609375\\
237	0.007242578125\\
238	0.007128671875\\
239	0.007265859375\\
240	0.007204921875\\
241	0.007259921875\\
242	0.007529921875\\
243	0.007401953125\\
244	0.007331484375\\
245	0.007432890625\\
246	0.00732046875\\
247	0.007263984375\\
248	0.007195546875\\
249	0.00732015625\\
250	0.007615859375\\
251	0.007217890625\\
252	0.00716546875\\
253	0.00741921875\\
254	0.007124453125\\
255	0.007366953125\\
256	0.00717609375\\
257	0.007129375\\
258	0.007275\\
259	0.007355546875\\
260	0.0074796875\\
261	0.007433984375\\
262	0.0073246875\\
263	0.007601875\\
264	0.00728765625\\
265	0.007312890625\\
266	0.0074353125\\
267	0.0072284375\\
268	0.007353515625\\
269	0.00751390625\\
270	0.007189375\\
271	0.0074659375\\
272	0.007199140625\\
273	0.007534921875\\
274	0.007476328125\\
275	0.007461953125\\
276	0.00714203125\\
277	0.00742625\\
278	0.007480390625\\
279	0.007228359375\\
280	0.007120625\\
281	0.00735734375\\
282	0.00733875\\
283	0.007519609375\\
284	0.007427421875\\
285	0.007518671875\\
286	0.00711265625\\
287	0.00737015625\\
288	0.007473046875\\
289	0.007406640625\\
290	0.007151953125\\
291	0.007368515625\\
292	0.007369140625\\
293	0.00755015625\\
294	0.007381171875\\
295	0.0073609375\\
};

\end{axis}
\end{tikzpicture}%
}
				\vspace{-0.6cm}	
				\caption{ \footnotesize BER evolution of code 1 at $\mathrm{SNR_{des}}=\unit[2]{dB}$.}	
				\vspace{0.25cm}	
				\label{fig:BP_epochs}
			\end{subfigure}
			
			\vspace{-0.1cm}
			
			\begin{subfigure}{\columnwidth}
				
				\resizebox{\columnwidth}{!}{
%
%
\definecolor{mycolor1}{rgb}{0.74902,0.74902,0.00000}%
\begin{tikzpicture}

\begin{axis}[%
width=\columnwidth,
height=0.225\columnwidth,
at={(2.516in,1.149in)},
scale only axis,
xmin=0,
xmax=150,
xlabel style={font=\color{white!15!black}},
xlabel={Population index},
ymode=log,
ymin=1e-3,
ymax=3.01e-3,
yminorticks=true,
ytick={1e-3, 2e-3, 3e-3},
yticklabels={1e-3, 2e-3, 3e-3},
xtick={0,50,100,150},
ylabel style={font=\color{white!15!black}},
ylabel={min(BER)},
axis background/.style={fill=white},
xmajorgrids,
xminorgrids,
ymajorgrids,
yminorgrids,
legend style={legend cell align=left, align=left, draw=white!15!black}
]
\addplot [color=mycolor1, line width=0.5pt]
  table[row sep=crcr]{%
1	0.003098125\\
2	0.00249640625\\
3	0.002552734375\\
4	0.00228859375\\
5	0.00184046875\\
6	0.00183875\\
7	0.001745390625\\
8	0.00172109375\\
9	0.001680078125\\
10	0.001697890625\\
11	0.0016028125\\
12	0.001537109375\\
13	0.001476796875\\
14	0.00141609375\\
15	0.001517578125\\
16	0.00149328125\\
17	0.0013809375\\
18	0.001493359375\\
19	0.001434609375\\
20	0.00143015625\\
21	0.00153875\\
22	0.001393359375\\
23	0.001402109375\\
24	0.001404296875\\
25	0.001374765625\\
26	0.001418046875\\
27	0.00142171875\\
28	0.001344609375\\
29	0.001416875\\
30	0.00135140625\\
31	0.001365703125\\
32	0.00133671875\\
33	0.001434921875\\
34	0.001353203125\\
35	0.001376484375\\
36	0.001239765625\\
37	0.0012625\\
38	0.001389453125\\
39	0.00127890625\\
40	0.001210703125\\
41	0.001389609375\\
42	0.0013140625\\
43	0.001217109375\\
44	0.001301875\\
45	0.001265234375\\
46	0.001237265625\\
47	0.00135640625\\
48	0.001369453125\\
49	0.001226015625\\
50	0.001294375\\
51	0.001334765625\\
52	0.001235625\\
53	0.0013340625\\
54	0.001316796875\\
55	0.0013275\\
56	0.00127640625\\
57	0.001281171875\\
58	0.001289453125\\
59	0.0012159375\\
60	0.00122484375\\
61	0.001294609375\\
62	0.0012040625\\
63	0.001158984375\\
64	0.001213125\\
65	0.001199453125\\
66	0.00119078125\\
67	0.00121015625\\
68	0.001211484375\\
69	0.00127328125\\
70	0.00123765625\\
71	0.001195\\
72	0.001210546875\\
73	0.001153203125\\
74	0.001241875\\
75	0.00117578125\\
76	0.00131328125\\
77	0.0012484375\\
78	0.001141875\\
79	0.001188671875\\
80	0.001170625\\
81	0.001231484375\\
82	0.0012403125\\
83	0.00129484375\\
84	0.001255390625\\
85	0.001216640625\\
86	0.00131\\
87	0.00124484375\\
88	0.001253828125\\
89	0.00119375\\
90	0.001266875\\
91	0.001211953125\\
92	0.00120109375\\
93	0.001198984375\\
94	0.001135078125\\
95	0.00125140625\\
96	0.001187265625\\
97	0.001208046875\\
98	0.00112640625\\
99	0.001226640625\\
100	0.0011525\\
101	0.001139375\\
102	0.0011215625\\
103	0.00106578125\\
104	0.001123828125\\
105	0.001163515625\\
106	0.00113984375\\
107	0.001176875\\
108	0.0010484375\\
109	0.001101015625\\
110	0.001154140625\\
111	0.00110609375\\
112	0.001186875\\
113	0.001143203125\\
114	0.0011790625\\
115	0.00112953125\\
116	0.00117515625\\
117	0.001144140625\\
118	0.001104765625\\
119	0.0011428125\\
120	0.001131640625\\
121	0.00118046875\\
122	0.001159140625\\
123	0.00113890625\\
124	0.001065234375\\
125	0.001115390625\\
126	0.001145546875\\
127	0.00114921875\\
128	0.001150234375\\
129	0.00117640625\\
130	0.0010709375\\
131	0.001138984375\\
132	0.001134921875\\
133	0.001159375\\
134	0.001130546875\\
135	0.001173046875\\
136	0.001155703125\\
137	0.0011271875\\
138	0.001179765625\\
139	0.001248515625\\
140	0.001141328125\\
141	0.001141640625\\
142	0.00111015625\\
143	0.001131171875\\
144	0.00117765625\\
145	0.00119921875\\
146	0.0011590625\\
147	0.001062109375\\
148	0.0011409375\\
149	0.001098671875\\
150	0.001078984375\\
151	0.001174296875\\
152	0.00113703125\\
153	0.001134921875\\
};

\end{axis}
\end{tikzpicture}%
}
				\vspace{-0.6cm}
				\caption{ \footnotesize BER evolution of code 2 at 
					$\mathrm{SNR_{des}}=\unit[2.5]{dB}$.}			
				\label{fig:BP_epochs2}
			\end{subfigure}
			\vspace{-0.0cm}\caption{\footnotesize Optimized $\mathcal{P}(512,256)$-codes  over the \ac{AWGN} 
				channel at 	$\mathrm{SNR_{des}}=\unit[2]{dB}$ and $\unit[2.5]{dB}$ 
				under 
				LDPC(-like) decoding and no CRC is used.}	 	%
			\vspace{-0.5cm}
			
		\end{figure}
		
				\begin{table}[H]
					\vspace{-0.25cm}
					\fontsize{9}{12}\selectfont 
					\begin{center}
						\caption{\footnotesize{The number of minimum-weight codewords of a 
								$\mathcal{P}$(256,128)-code}} \vspace{-0.1cm}
						\label{tab:A_dmin}
						\begin{tabular}{ccc}
							\hline 
							Construction  & $d_{min}$ & $A_8$\tabularnewline
							\hline 
							$5$G \cite{polar5G2018} & 8 & 96 \tabularnewline
							
							Optimized @ $\unit[3]{dB}$ & 8 & 68 \tabularnewline
							\hline 
						\end{tabular}
					\end{center}	
					\vspace{-0.6cm}
				\end{table}
		
		\vspace{0.1cm}
		
		\subsection{Codeword length $N=512$}

		A $\mathcal{P}(512,256)$ polar code along with its corresponding sparse $\mathbf{H}_{\text{pruned}}$ are designed under LDPC-like decoding with BER as the cost function. The newly obtained code is compared to the 5G polar code without the CRC-aid \cite{polar5G2018}, under BP, LDPC-like and plain  \ac{SCL} decoding. As depicted in  Fig. \ref{fig:BER512}, BER performance of the optimized code under LDPC(-like) decoding achieve the performance of the 5G polar code  under Arıkan's conventional BP decoder, with a performance gain of around $0.4$ dB at BER of $10^{-3}$, compared to the originally proposed code in \cite{sparseBP}. It even outperforms  the 5G polar code  under  Arıkan's conventional BP decoder, while approaching it under plain \ac{SCL} in the high SNR range.

		For this scenario, two optimization processes were conducted, at design SNRs 2.5 dB and 3 dB, resulting in code 1 and code 2, respectively. Depending on the desired SNR region (or error-rate level), one of them may be more suited to be applied. Furthermore, the convergence behavior (or epochs) of both processes is depicted in Fig. \ref{fig:BP_epochs} and \ref{fig:BP_epochs2}, respectively.
		
		
		It is worth-mentioning that the CRC-aided 5G polar codes under SCL decoding are of better error-rate performance than the state-of-the-art BP-based polar decoders. One reason is the CRC incompatibility in the iterative decoding environment.
		
		\section{Conclusion} \label{sec:conc}
		An enhanced polar code design tailored to off-the-shelf LDPC decoders is presented. 
		We attempt to design a suitable parity-check matrix such that polar codes could be decoded using an equivalent LDPC-like code. 
		Thus, our proposed codes can be efficiently encoded with a low complexity polar encoder and can be decoded with a low complexity conventional (flooding schedule-based) BP LDPC decoder, also providing soft-outputs.
		Extensions to other LDPC decoders (e.g., min-sum approximation-based or quantized LDPC decoders) are straightforward.
		Using our proposed design method, error-rate performance gains were achieved compared to the 5G polar codes under iterative decoding. We further showed that the gains can be attributed to the reduction in the number of minimum-weight codewords and other enhancements such as girth profile of the respective decoding graph and the reduced number of punctured variable nodes in it.

		\vspace{-0.15cm}
		
		
		\bibliographystyle{IEEEtran}
		\bibliography{references}
	\end{NoHyper}
\end{document}